\documentclass[prb,twocolumn,nofootinbib,citeautoscript,eqsecnum,10pt]{revtex4-1}
\pdfoutput=1

\usepackage{dsfont}
\usepackage{float}
\usepackage{array}
\usepackage{slashed,bbold}

\usepackage{amsmath,amssymb,bm} 
\usepackage{graphicx}

\usepackage[tight]{subfigure} 

\usepackage{color} 
\usepackage[papersize={8.5in,11in}]{geometry}

\newcommand{\bk}{{\bf k}}

\newcommand{\bp}{{\bf p}}

\def \be{\begin{equation}}
\def \ee{\end{equation}}
\def \bea{\begin{eqnarray}}
\def \eea{\end{eqnarray}}
\def \nn{\nonumber \\}

\begin{document}

\title{Effect of interactions on the quantization of the chiral photocurrent for double-Weyl semimetals}

\author{Ipsita Mandal}

\affiliation{Faculty of Science and Technology, University of Stavanger, 4036 Stavanger, Norway}

\begin{abstract}
{The circular photogalvanic effect (CPGE) is the photocurrent generated in an optically active material in response to an applied ac electric field, and it changes sign depending on the chirality of the incident circularly polarized light. It is a non-linear dc current as it is second-order in the applied electric field, and for a certain range of low frequencies, takes on a quantized value proportional to the topological charge for a system which is a source of nonzero Berry flux. We show that for a non-interacting double-Weyl node, the CPGE is proportional to two quanta of Berry flux. On examining the effect of short-ranged Hubbard interactions upto first-order corrections, we find that this quantization is destroyed. This implies that unlike the quantum Hall effect in gapped phases or the chiral anomaly in field theories, the quantization of the CPGE in topological semimetals is not protected.}
\end{abstract}
\maketitle

\tableofcontents

\section{Introduction}
Semimetals are materials which can support gapless quasiparticle excitations in two or three dimensions, in the vicinity of isolated band touching points in the Brillouin zone, thus possessing discrete Fermi points (rather than Fermi surfaces). They come in different varieties, for example, the Fermi points may appear at linear band crossings ({\it e.g.} graphene, Weyl semimetals), or at quadratic band crossings \cite{abrikosov,balents-moon} ({\it e.g.} Luttinger semimetals). A more non-trivial example of such semimetal is the double-Weyl semimetal, which consists of two bands touching each other linearly along one momentum direction, but quadratically along the remaining directions \cite{pardo,pardo2,montambaux,hasegawa,kush-ips}. Some of these three-dimensional (3d) semimetals ({\it e.g.} Weyl and double-Weyl semimetals) possess a nonzero Berry curvature at the Fermi nodes. In this paper, we focus on the 3d double-Weyl semimetals \cite{balents-moon,fang,bernevig}, which, in the momentum space, have double the monopole charge of Weyl semimetals.

A double-Weyl semimetal can be realized by applying a Zeeman field to an isotropic Luttinger semimetal \cite{balents-moon}. They are also predicted to appear \cite{bernevig,Huang1180,fang} in SrSi$_2$, and in the ferromagnetic phase of HgCr$_2$Se$_4$. Our aim is to study the circular photogalvanic effect (CPGE), also known as chiral photocurrent. The CPGE refers to
the dc current, that is generated as a result of shining circularly polarized light on the surface of an optically active  metal \cite{claudio,sipe,joel-prl,nastos}. In fact, the CPGE refers to the part of the photocurrent that switches sign with the sign of the helicity of the incident polarized light. This is a non-linear response, as it is second order in the applied ac electric field, and at low frequencies, it depends on the orbital Berry phase of the Bloch electrons. Hence, CPGE is a measure of the topological charge at a Fermi node possessing a nontrivial Berry curvature. 

The quantization of the CPGE has been demonstrated in earlier works for the topological Weyl nodes \cite{grushin,nagaosa}. In this paper, we will consider the issue of quantization of CPGE for the double-Weyl nodes.
Firstly, we will show that in the absence of interactions, the CPGE is indeed proportional to the topological charge of the node at low enough frequencies. Secondly, we will examine the effect of Hubbard interactions on this quantized value.


\section{The continuum Hamiltonian for a double-Weyl semimetal}

The Hamiltonians describing a pair of double-Weyl nodes can be written in the form \cite{balents-moon,fang,bernevig}
\begin{equation}
  \label{eqham}
  \mathcal{H}_\pm = \mathbf{b}_\pm(\mathbf{k}) \cdot \mathbf\sigma,
\end{equation}
with
\begin{equation}
  \label{eqham1}
  \mathbf{b}_\pm(\mathbf{k}) = \left( \begin{array}{c} -\frac{\sqrt{3}}{2}\left (k_x^2-k_y^2\right ) \\
    \sqrt{3} \,k_x\, k_y \\ \mp v\, k_z \end{array}\right) .
\end{equation}
Here, $\sigma_{i}$ $\left (i= x, y, z\right )$ are the three Pauli matrices, and the ``$\pm$'' sign reflects the two opposite chiralities of the two nodes.
The energy eigenvalues are:
\begin{eqnarray}
E_{\pm}(\mathbf k)=\pm \sqrt{v^2 \,k_z^2 + \frac{3}{4}(k_x^2+ k_y^2)^2}\,.
\end{eqnarray}

For each the given two-band Hamiltonians, we can define an $U(1)$
Berry curvature, which is analogous to a magnetic field in momentum space. This Berry curvature is given by:
\begin{equation}
  \label{eqberry}
  \mathcal{B}_\pm^i = \frac{1}{8\pi}\,\epsilon^{ijl}\, \hat{b}_\pm
  \cdot \partial_{ k_j} \hat{b}_\pm \times \partial_{k_l} \hat{b}_\pm\,,
\end{equation}
where $\hat{b}_\pm = \mathbf{b}_\pm/|\mathbf{b}_\pm|$. It is easy to check that this magnetic field is
divergenceless $\left( \partial_{k_j} \mathcal{B}_\pm^j=0\right) ,$ as long as it is computed in regions away from the points of
singularity where $ \mathbf{b}_\pm =0$.  The band touching
point is such a singularity, where we have:
\begin{equation}
  \label{eqdiv}
  \partial_{k_j} \mathcal{B}^j_{\pm}(\mathbf{k}) = \pm 2\, \delta(\mathbf{k})\,.
\end{equation}
Thus each double-Weyl node is a source of two Berry flux quanta. These nodes come in pairs, sourcing 
equal and opposite flux quanta, such that the sum of
Berry flux quanta from both the double-Weyl nodes vanishes, which
is the desired physical scenario as the Brillouin zone is a closed manifold without any
boundary through which no net flux can emanate.

\section{Quantization of CPGE in the absence of interactions}
 \label{non-interacting}
 
The CPGE tensor is defined as \cite{grushin,nagaosa}:
\begin{align}
\beta_\pm^{ij} &=\frac{\mathrm{i} \, \pi\,e_A^3}  {h^2}
\int d^3k \left[ \partial_{k_i} \left( E_{+} - E_{-} \right) \right]
\mathcal{B}_{\pm}^j \,\delta\left( \hbar \,\omega - E_{+} + E_{-} \right),
\end{align}
where $e_A$ is the electric charge.
To perform the integrals, we change variables as follows:
\begin{align}
& k_r =  \sqrt{\mathcal{R} \, \sin \theta }\,,
\quad k_z =\frac{\sqrt{3} \,\mathcal{R} \cos \theta } {2 \,v}\,,\nonumber \\
& k_x = k_r \cos \phi \,, \quad
k_y =  k_r  \sin \phi \,, \nonumber \\
&\text{ where }
0 \leq \mathcal{R} \leq \infty\,,\,\,
0 \leq \theta \leq \pi \text{ and }
0 \leq \phi \leq 2\,\pi \,.
\end{align}

Using the above, we get:
\begin{align}
\beta_\pm^{11} =\beta_\pm^{22}=\beta_\pm^{33}= \left(  \pm 2 \right) \times \frac{\mathrm{i} \, \pi\,e_A^3}  {3\,h^2} \,.
\end{align}
All non-diagonal components $\left( \beta_\pm^{ij} \big \vert_{i \neq j}\right)$ evaluate to zero. Clearly, we see that
\begin{align}
\text{tr}[\beta_\pm] = \left(  \pm 2 \right) \times \frac{\mathrm{i} \, \pi\,e_A^3}  {h^2} \,,
\end{align}
where is $\pm 2 $ the monopole charge of the corresponding double-Weyl node.
The time derivative of the injection current is defined as the second order response
\begin{align}
\frac{dj_i^\pm}{dt} = \beta_\pm^{ij} \left[ \mathbf{E} (\omega) \times  \mathbf{E}^* (\omega) \right ]_j\,,
\label{cpgej}
\end{align}
to an electric field $  \mathbf{E} (\omega) = \mathbf{E}^* (-\omega)$. Therefore, the CPGE is also quantized.

Now let us compute the second-order photocurrent from the field-theoretic definition, using Feynman diagrams.
Firstly, we need the three components of the paramagnetic current operator (using $ \mathcal{J}_i^\pm  (\mathbf k) \equiv  e_A\, \frac{\delta\mathcal{H}_\pm (\mathbf k)}{\delta k_i}$), which are given by:
\begin{align}
& \mathcal{J}_x(\mathbf k)
  =  e_A\, \sqrt{3} \left(- k_x \,\sigma_x  + k_y\,\sigma_y  \right) \,,\nonumber \\
& \mathcal{J}_y(\mathbf k)  = e_A\,  \sqrt 3 \left( k_y \,\sigma_x  + k_x\,\sigma_y  \right)  \,,\nonumber \\
& \mathcal{J}_z(\mathbf k)^\pm = \mp   e_A\, v\, \sigma_z \,.
\end{align}
From now on we will drop the ``$\pm$" subscript/superscript and concentrate only on the double-Weyl node with charge $+2$, unless stated otherwise. This is justified when the dc contribution to the photocurrent can be calculated separately for
each node, such as when the nodes are well separated in the momentum space.

\begin{figure}[htb]
{\includegraphics[width = 0.15 \textwidth]{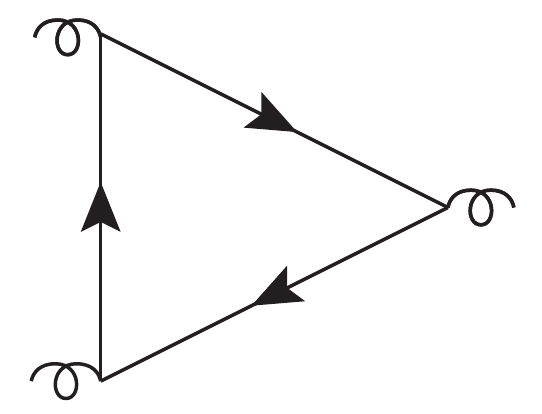}} 
\caption{Feynman diagram contributing to the quantized circular photogalvanic effect in the absence of interactions.}
\label{fig-bareresponse}
\end{figure}

The expression for the second-order photocurrent is given by:
\begin{align}
j_i(\Omega) &= -\frac{ \chi_1^{ jli}(\omega_1, \omega_2)
 +  \chi_2^{jli}(\omega_1, \omega_2)} {\hbar^2} \,A^{j} (\omega_1)\, A^{l}(\omega_2) 
 \nonumber \\
 &= \frac{ \chi_1^{ jli}(\omega_1, \omega_2)
 +  \chi_2^{jli}(\omega_1, \omega_2)} {\hbar^2\,\omega_1 \, \omega_2} \,E^{j} (\omega_1)\, E^{l}(\omega_2) \,, 
\label{currentgeneral}
\end{align}
where $\Omega \equiv \omega_1 + \omega_2,$ and the contributions $\chi_1^{ jli}$ and $\chi_2^{jli}$ are given by Feynman diagrams of the type shown in Fig.~\ref{fig-bareresponse}. In the second line, we have used the relation between the electric field and the vector potential, which is: $ \mathbf{E} (\omega) = \mathrm{i} \, \omega\,\mathbf{A} (\omega)$. 

\begin{widetext}
We compute the analytical expressions for $\chi_{1,2}^{ jli}$ in the Matsubara formalism, such that
\begin{align}
\chi^{ijl}_1(\mathrm{i} \,\omega_1, \mathrm{i} \, \omega_2) 
= T\sum_{\varepsilon_n} \int \frac{d^3 k}{(2\pi)^3} \text{tr} \left[ \mathcal{J}_i \,G(\mathrm{i} \,\varepsilon_n - \mathrm{i} \,\omega_1, \bk)\, \mathcal{J}_j\, G(\mathrm{i} \,\varepsilon_n 
- \mathrm{i} \, \Omega, \bk)\, \mathcal{J}_l\,G(\mathrm{i} \,\varepsilon_n, \bk)\right],
\end{align}
where $T$ is the temperature, $n$ is an integer, and $\varepsilon_n = \left (2\,n+1 \right) \pi \, T$.
\end{widetext} 
In the zero temperature limit, we can use $T \sum \limits_{\varepsilon_n}\ldots \to \int \frac{d\varepsilon} {2\,\pi} \ldots\,.$ 
Furthermore, from the expression for $ \chi^{ijl}_1(\mathrm{i} \,\omega_1, \mathrm{i} \, \omega_2)  $, we can obtain $\chi_{2}^{ jli}$ by using the relation:
\begin{align}
\chi^{ijl}_2( \mathrm{i} \,  \omega_1 , \mathrm{i} \, \omega_2) = \chi^{jil}_1( \mathrm{i} \,  \omega_2, \mathrm{i} \, \omega_1)\,.
\label{eqchi12}
\end{align}

In the absence of interactions, we can calculate the contributions from each node separately. The Green's function for the first double-Weyl node is given by: 
\begin{align}
G(\mathrm{i} \,\varepsilon_n , \mathbf k) = \frac{1}{2} 
\left[\frac{ \mathbb{1} + \hat{b}_+ (\mathbf k)\cdot \mathbf{\sigma}}
{ \mathrm{i} \, \varepsilon_n -E_+(\mathbf{k}) -|\mu|} 
+ \frac{ \mathbb{1} - \hat{b}_+ (\mathbf k) \cdot \mathbf{\sigma}} 
{\mathrm{i} \,\varepsilon_n + E_+(\mathbf{k}) -|\mu|} \right],
\end{align}
where we have introduced the projectors $\left( \mathbb{1} \pm \hat{b}_+ (\mathbf k)\cdot \mathbf{\sigma} \right)$ onto the conduction (``+'') and the valence (``-'') bands, and have chosen the chemical potential $\mu $ to be negative for definiteness ({\it i.e.} $\mu < 0$).
Similarly, the Green's function for the second double-Weyl node is given by: 
\begin{align}
\tilde G(\mathrm{i} \,\varepsilon_n , \bk) = \frac12 \left[\frac{\mathbb{1} + \hat{b}_- (\mathbf k)\cdot \mathbf{\sigma}}
{ \mathrm{i} \, \varepsilon_n - E_+(\mathbf{k}) +|\tilde \mu|} 
+ \frac{ \mathbb{1} - \hat{b}_- (\mathbf k) \cdot \mathbf{\sigma}} {\mathrm{i} \,\varepsilon_n 
+ E_+(\mathbf{k}) + |\tilde \mu|} \right],
\end{align}
where we have chosen $\tilde \mu> 0 $ for definiteness.

\begin{widetext}
Performing all the integrals, we finally get:
\begin{align}
& \chi^{123}_1(\mathrm{i} \,\omega_1, \mathrm{i} \, \omega_2) =  \int \frac{d\varepsilon \,d^3 k}{(2\,\pi)^4} \text{tr}  \left[ \mathcal{J}_x \,G(\mathrm{i} \,\varepsilon_n - \mathrm{i} \,\omega_1, \bk)\, \mathcal{J}_y\, G(\mathrm{i} \,\varepsilon_n 
- \mathrm{i} \, \Omega, \bk)\, \mathcal{J}_z\,G(\mathrm{i} \,\varepsilon_n, \bk)\right] \nonumber\\
& =
\frac{e_A^3 \left [ \omega_1^3 \left ( \omega_1+2 \,\omega_2 \right ) \ln \left(4 \,\mu^2
+ \omega_1^2\right)- \omega_2^3 (2 \,\omega_1+\omega_2)\, \ln \left(4 \,\mu^2+\omega_2^2\right)+(\omega_2-\omega_1) (\omega_1+\omega_2)^3 \, \ln \left(4 \,\mu^2+(\omega_1+\omega_2)^2\right) \right ]}
{24 \,\pi ^2 \, \omega_1 \, \omega_2 \left (\omega_1+\omega_2 \right )}
\end{align}
for $T \rightarrow 0 \,.$
\end{widetext}
One can check that $\chi^{ijl}_1 \propto \varepsilon_{ijl}$, and hence the computation of $ \chi^{123}_1$ is sufficient to know all the nonzero components of $ \chi^{ijl}_1  $.

We need to find the physical response through the analytical continuation of the above expressions from Matsubara frequencies to real frequencies. This is a subtle procedure which should be carried out carefully. Choosing $\omega_{1,2} > 0$ for definiteness, the analytical continuation is  performed by taking \cite{lopes,kozii}
\begin{align}
\mathrm{i}\, \omega_{1,2} \to \omega_{1,2} + \mathrm{i}\,\delta\,, \qquad \delta \to +0\,. 
\label{analyticalcontinuation}
\end{align}
The logarithms then transform according to
\begin{align}
& \ln\left[4\,\mu^2 + \omega^2 \right] 
\nonumber \\ &
\to \ln\left[4\,\mu^2 - (\omega + \mathrm{i}\, \delta)^2 \right]
\nonumber \\ &
\qquad  = \ln |4\,\mu^2 - \omega^2|- \mathrm{i} \, \pi\,  \text{sign}( \omega) \,  \Theta\big(|\omega| - 2\,|\mu|\big) \,.
\end{align}
We then need to set $ \omega_1 =  \Omega- \omega_ 2  $ with $\Omega \rightarrow 0$.
After the analytical continuation, we find that in this limit,
\begin{align}
\chi_1^{123}( \omega + \Omega, -\omega) 
\overset{\Omega \rightarrow 0 } {=}
  -\frac{e_A^3\, \omega^2}{ 12 \,\pi\,\Omega}  \, \Theta \big(\omega - 2\,|\mu| \big ) \,. \label{chi1answer}
\end{align}
An identical contribution comes from 
$\chi_2^{123}$ (on using Eq.~(\ref{eqchi12})). Adding these together, we find that the current expression  in Eq.~(\ref{currentgeneral}) reduces to:
\begin{align}
j_l =  \frac{2\,\pi\,e_A^3} {3\,h^2\,\Omega}\, \varepsilon_{ijl} \, E^{i} (\omega + \Omega) \,E^{j}(-\omega) 
\,\Theta\big(\omega - 2\,|\mu |\big) \,.
\end{align}
In the time domain, this corresponds to
\begin{align}
& \frac{d j_i} {dt} = \frac{\mathrm{i} \, \beta_0(\omega)} {3}
 \left[ \mathbf{E}(\omega) \times \mathbf{E}(-\omega)  \right]_{i} \,,
 \nonumber \\
 & \beta_0(\omega) \equiv \frac{2\,\pi\,e_A^3 \,\Theta\big(\omega - 2\,|\mu|\big)} { h^2} \,. 
 \label{djbydt}
\end{align}
This agrees with Eq.~(\ref{cpgej}).

\begin{figure}[htb]
\subfigure[]{\includegraphics[width = 0.14 \textwidth]{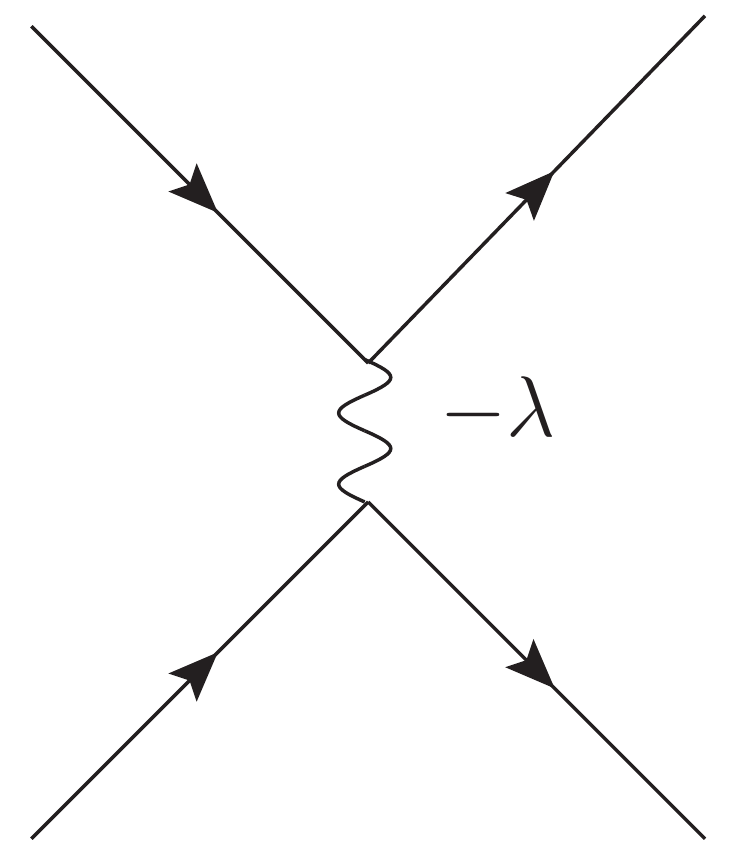}} \quad
\subfigure[]{\includegraphics[width = 0.14 \textwidth]{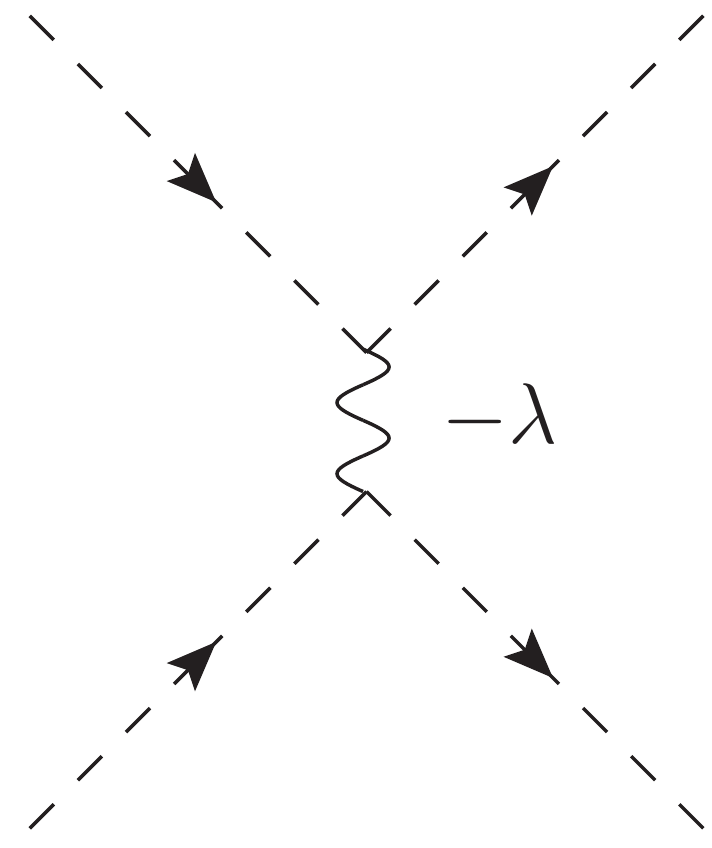}}\quad
\subfigure[]{\includegraphics[width = 0.14 \textwidth]{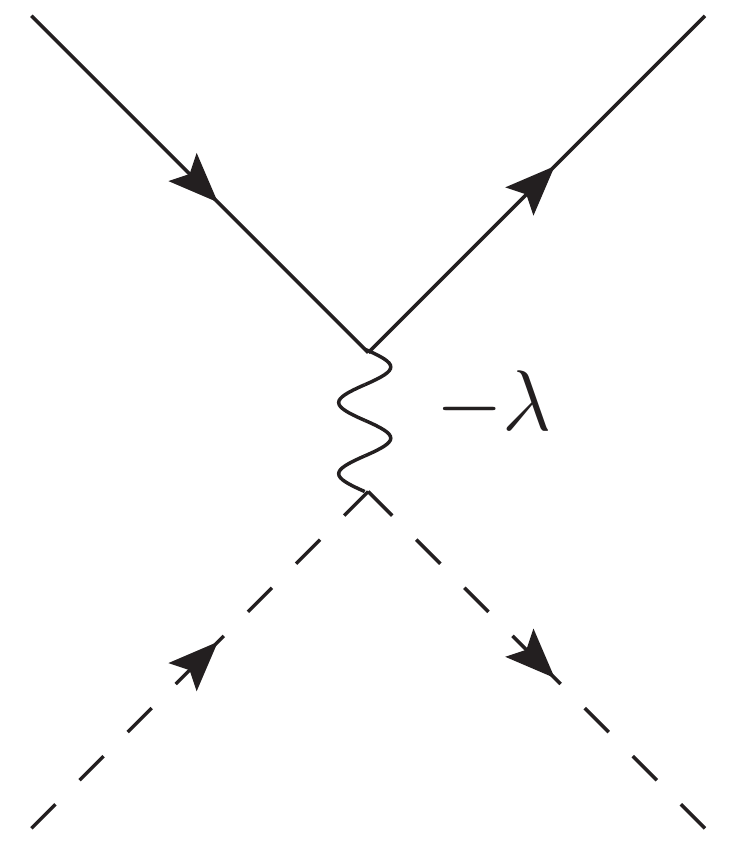}} 
\subfigure[]{\includegraphics[width = 0.14 \textwidth]{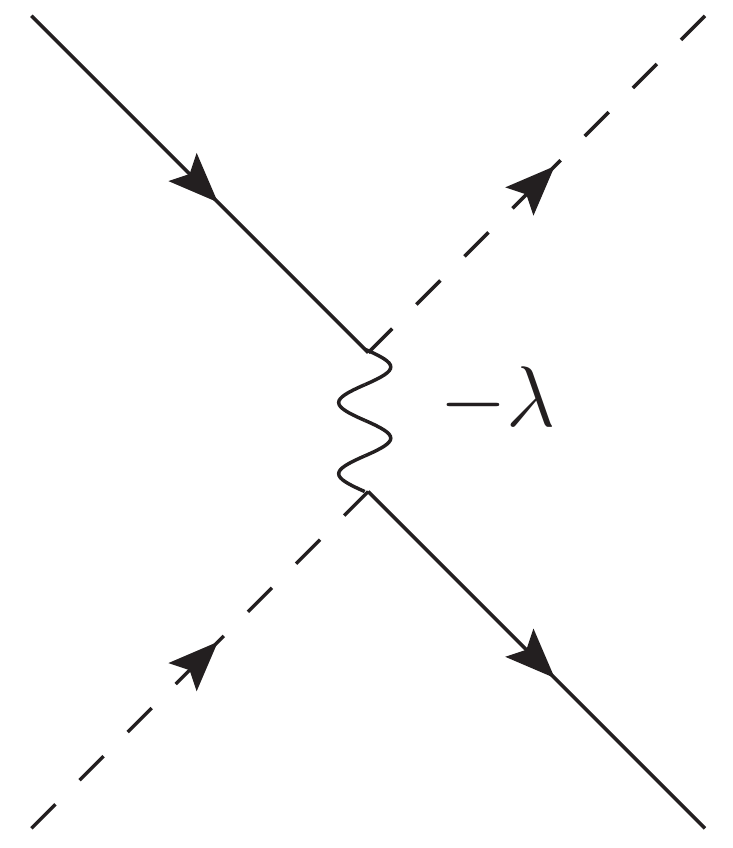}}\quad
\subfigure[]{\includegraphics[width = 0.14 \textwidth]{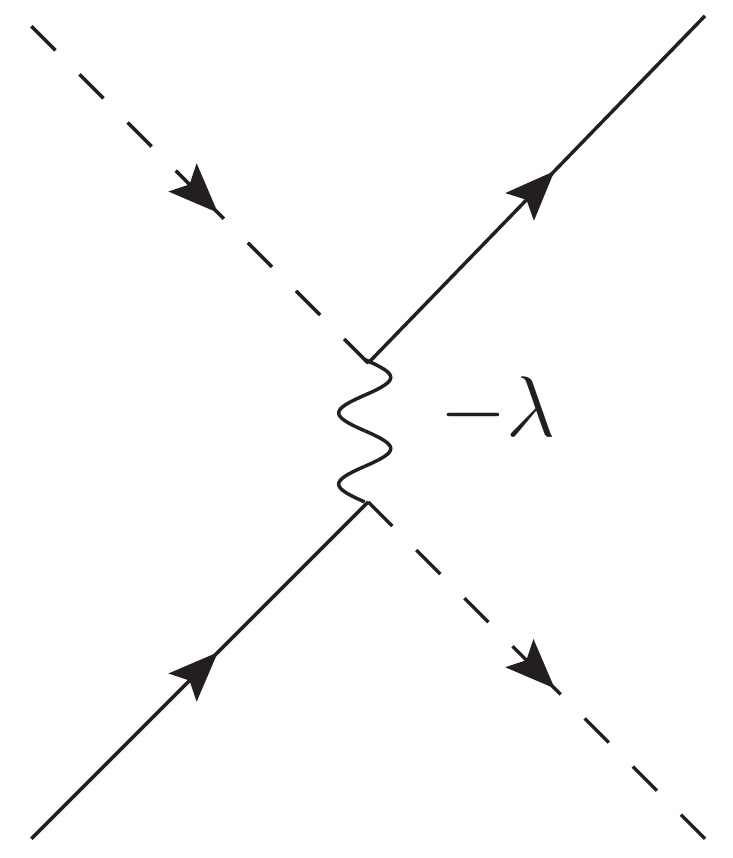}}
\caption{Feynman diagrams contributing to the scattering processes for Hubbard interactions, described by Eq.~(\ref{Hintgeneral}). Here, a solid line represents the Green's function of the first node (with chemical potential $\mu$), while a dashed line represents the Green's function of the second node (with chemical potential $\tilde \mu$). The wavy lines represent the four-fermion interactions. Hence, diagrams (a)-(c) involve only intranodal scatterings, whereas (d)-(e) describe internodal processes.}
\label{Fig:scattering}
\end{figure}

This result from the non-interacting case has been obtained for the first double-Weyl node with the chemical potential $\mu$. Analogously, for the second node, we would obtain:
\begin{align}
\tilde \beta_{0}(\omega) = - \frac{2\,e_A^3 \,\Theta\big(\omega - 2\,|\tilde \mu|\big)} { \pi\,h^2} \, . 
\end{align}
Consequently, in the frequency range $2\,|\mu| < \omega < 2 \,|\tilde \mu|$, only the first node contributes to the CPGE, while the contribution from the second node is zero due to Pauli blocking.


\section{Corrections to the quantized CPGE due to short-ranged Hubbard interactions} 
\label{sechubbard}

In this section, we consider the first-order perturbative corrections originating from four-fermion interactions. The interaction Hamiltonian for short-ranged Hubbard interactions is given by:
\begin{align}
&H_{\text{int}} \nonumber \\
=& \frac{-\lambda}{2} 
\sum \limits _{s, s'}
\int \frac{d^3 k \, d^3 p}{(2\pi)^6} 
\Big [ \sum \limits _{\zeta ,\eta =1}^{2} \psi^\dagger_{ \zeta,s}(\mathbf k)\,
 \psi_{\zeta,s}(\mathbf k)\, \psi^\dagger_{\eta,s'}(\mathbf p) \psi_{\eta, s'}(\mathbf p) \nonumber \\
& \hspace{2.7 cm}
+ \sum \limits _{\zeta  =1}^{2} 
\psi^\dagger_{\zeta,s}(\mathbf k)\, \psi_{\bar \zeta,s}(\bk)\, \psi^\dagger_{\bar \zeta,s'}(\mathbf p)
\, \psi_{\zeta,s'}(\bp) \Big] \, ,
\label{Hintgeneral}
\end{align}
where $\lambda $ is the Hubbard interaction strength (positive $\lambda$ corresponds to the attractive interaction), and $\psi_{\zeta,s}(\mathbf k)$ denotes the fermion field with nodal index $\zeta$ and pseudospin index $s$. The first and the second terms describe the intranodal and internodal scattering processes respectively. These are shown diagrammatically in Fig.~\ref{Fig:scattering}. In the diagrams, we have used a solid line to represent the Green's function for the first double-Weyl node, and a dashed line to depict the Green's function for the second double-Weyl node. In the following subsections, we will compute the first-order self-energy and vertex corrections due to the Hubbard interactions.

\subsection{First-order self-energy corrections}

\begin{figure}[htb]
\subfigure[]{\includegraphics[width = 0.225 \textwidth]{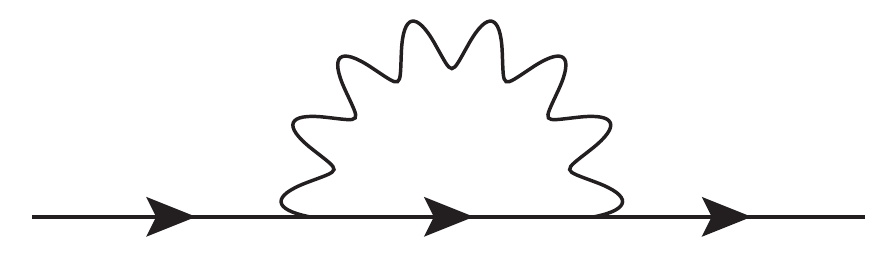}} \hspace{0.3 cm}
\subfigure[]{\includegraphics[width = 0.225 \textwidth]{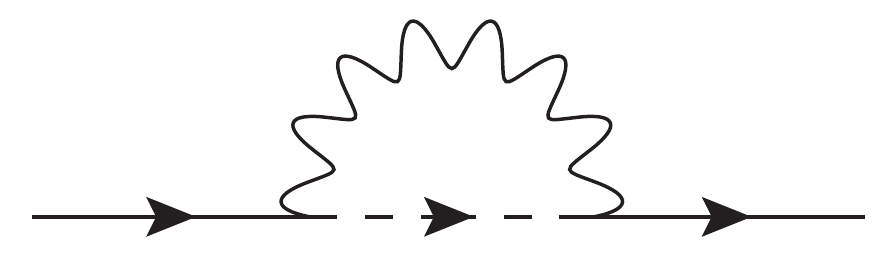}}
\subfigure[]{\includegraphics[width = 0.1 \textwidth]{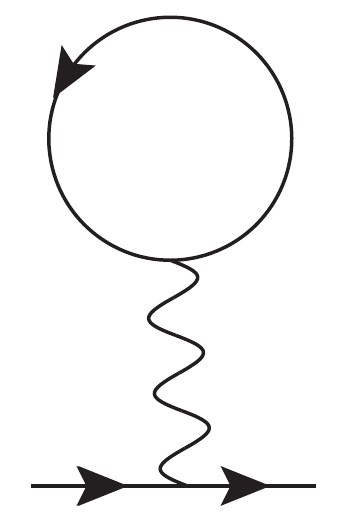}} \hspace{2.5 cm}
\subfigure[]{\includegraphics[width = 0.1 \textwidth]{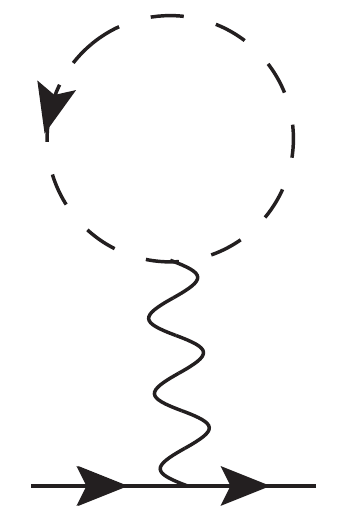}}
\caption{Feynman diagrams contributing to first-order corrections to self-energy. Diagrams (a) and (c) depict the internodal scatterings, while diagrams (b) and (d) describe the internodal scatterings.}
\label{Fig:self-energy}
\end{figure}

The contributions to the first-order self-energy correction are given by the Feynman diagrams shown in Fig.~\ref{Fig:self-energy}. For the short-ranged Hubbard interaction, scatterings between double-Weyl nodes of opposite chiralities have to be taken into account, which are given by the second term of Eq.~(\ref{Hintgeneral}). The analytic expression for Fig.~\ref{Fig:self-energy}(a) reads as:
\begin{align}
\Sigma^{(a)} &= \lambda \,T \sum \limits_{\varepsilon_n}\int \frac{ d^3 k}{(2\,\pi)^3}\, 
G(\mathsf{i}\, \varepsilon_n, \mathbf k)
\nonumber \\
& \overset{T \rightarrow 0 } {=} -\frac{\lambda}2 
\int \frac{d^3 k}{(2\pi)^3} \left[ 1 - \Theta( E_+ - |\mu|) \right] = - \frac{\lambda \,N_h} {2} \,,
\end{align}
where $N_h > 0$ is the number of holes below the double-Weyl point in the first node.
In a similar fashion, the contribution from Fig.~\ref{Fig:self-energy}(b) evaluates to:
\begin{align}
\Sigma^{(b)} &= \lambda \,T \sum \limits_{\varepsilon_n}\int \frac{ d^3 k}{(2\,\pi)^3}\, \tilde G(\mathsf{i}\, \varepsilon_n, \mathbf k)
 = \frac{\lambda \, N_e}{2} \,,
\end{align}
with $N_e > 0$ denoting the number of electrons above the double-Weyl point in the second node.

Finally, the contributions from Figs.~\ref{Fig:self-energy}(c) and \ref{Fig:self-energy}(d) evaluate to:
\begin{align}
\Sigma^{(c)} &= -\lambda \,T \sum \limits_{\varepsilon_n}\int \frac{ d^3 k}{(2\,\pi)^3}\, 
\text{tr} \left[G(\mathsf{i}\, \varepsilon_n, \mathbf k) \right] = -2 \,\Sigma^{(a)}\,, \nonumber \\
 \Sigma^{(d)} & = -\lambda \,T \sum \limits_{\varepsilon_n}\int \frac{ d^3 k}{(2\,\pi)^3}\, 
\text{tr} \left[\tilde G(\mathsf{i}\, \varepsilon_n, \mathbf k) \right]= -2 \,\Sigma^{(b)}\,,
\end{align}
resulting in the total self-energy
\begin{align}
\Sigma = \Sigma^{(a)} + \Sigma^{(b)} + \Sigma^{(c)} + \Sigma^{(d)}
 = -\frac{ \lambda \left( N_e - N_h  \right)} {2} \,.
\end{align}
The effect of this self-energy is to simply shift the chemical potential by an amount
\begin{align}
\delta \mu = - \Sigma = \frac{\lambda \left( N_e - N_h  \right)}{2} \,. 
\label{deltamu}
\end{align}
Clearly, this does not change the CPGE current, as it only modifies the frequency range where the quantized value of the CPGE is valid.

\subsection{First-order vertex corrections}

\begin{figure}[htb]
\subfigure[]{\includegraphics[width = 0.1 \textwidth]{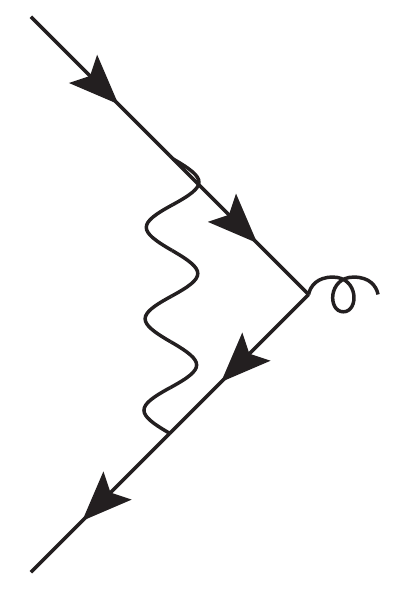}} \hspace{1 cm}
\subfigure[]{\includegraphics[width = 0.1 \textwidth]{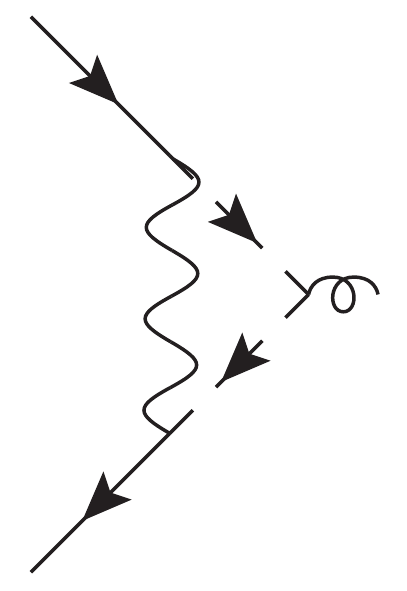}}
\caption{Feynman diagrams contributing to first-order vertex corrections. Diagrams (a) and (b) describe the intranodal and internodal scatterings, respectively.}
\label{figvertex}
\end{figure}

The Feynman diagrams contributing to first-order vertex corrections are shown in Fig.~\ref{figvertex}.
When the vertex $i=x$, with the external Matsubara frequency set to $\omega_1$ for definiteness, Fig.~\ref{figvertex}(a) contributes as:
\begin{align}
&\sqrt{3} \left(- k_x \,\sigma_x  + k_y\,\sigma_y  \right)  \nonumber \\
& \rightarrow   \lambda \sqrt{3}  \int \frac{d\varepsilon \,d^3 k}{(2\,\pi)^4} \,
   G(\mathrm{i}\, \varepsilon, \mathbf{k}) \left( k_y\,\sigma_y - k_x \,\sigma_x  \right) 
 G( \mathrm{i}\, \varepsilon - \mathrm{i}\,\omega_1, \mathbf{k}) 
\nonumber \\ & = 0\,.
\end{align}
Similarly, for $i=y$, Fig.~\ref{figvertex}(a) gives:
\begin{align}
&\sqrt{3} \left( k_y \,\sigma_x  + k_x\,\sigma_y  \right)  \nonumber \\
& \rightarrow   \lambda \sqrt{3}  \int \frac{d\varepsilon \,d^3 k}{(2\,\pi)^4} \,
   G(\mathrm{i}\, \varepsilon, \mathbf{k}) \left( k_y \,\sigma_x  + k_x\,\sigma_y  \right) 
 G( \mathrm{i}\, \varepsilon - \mathrm{i}\,\omega_1, \mathbf{k})
 \nonumber \\
 & = 0\,.
\end{align}
The only non-vanishing contribution from Fig.~\ref{figvertex}(a) comes for $i=z$, which gives:
\begin{align}
& -v\,\sigma_z   \nonumber \\
& \rightarrow - \lambda \,v \int \frac{d\varepsilon \,d^3 k}{(2\,\pi)^4} \,
   G(\mathrm{i}\, \varepsilon, \mathbf{k}) \,\sigma_z\,
 G( \mathrm{i}\, \varepsilon - \mathrm{i}\,\omega_1, \mathbf{k}) \nonumber \\
&  = \lambda \frac{6\, \Lambda -\sqrt{3} \left[4 \, |\mu|+ \mathrm{i}\, \omega_1 
  \ln \left(-\frac{\left(\sqrt{3} \Lambda 
  + \mathrm{i}\, \omega_1\right) (\omega_1+2 \,\mathrm{i}\,  |\mu|)} {\left(\sqrt{3} \Lambda 
  -\mathrm{i}\, \omega_1\right) (\omega_1-2 \,\mathrm{i}\, |\mu| )}\right)\right ]\sigma_z}
  {192 \,\pi }\,,
\end{align}  
where $\Lambda$ is the UV momentum cutoff.

The contribution from the diagram in Fig.~\ref{figvertex}(b) is analogous, but has an overall opposite sign due to the opposite chirality of the second node, and with $ |\mu| \rightarrow |\tilde \mu|$:
\begin{align}
& v\,\sigma_z  \nonumber \\
& \rightarrow
 \lambda \,v \int \frac{d\varepsilon \,d^3 k}{(2\,\pi)^4} \,
  \tilde  G(\mathrm{i}\, \varepsilon, \mathbf{k}) \,\sigma_z\,
\tilde  G( \mathrm{i}\, \varepsilon - \mathrm{i}\,\omega_1, \mathbf{k}) \nonumber \\
& =\lambda \frac{-6 \,\Lambda + \sqrt{3} \left[4 \, |\tilde \mu| - \mathrm{i}\, \omega_1 
\ln \left(-\frac{\left(\sqrt{3} \Lambda  + \mathrm{i}\, \omega_1\right) 
(\omega_1+2 \,\mathrm{i}\,  |\tilde \mu|)} 
  {\left(\sqrt{3} \Lambda -\mathrm{i}\, \omega_1\right) (\omega_1-2 \,\mathrm{i}\, | \tilde \mu| )}\right)\right ]\sigma_z}
  {192 \,\pi }\,.
\end{align} 
Adding these two contributions together, we find that for the first node, the vertex with $\sigma_z$ (and external frequency $\omega_1$) is renormalized according to:
\begin{align}
& -v\,\sigma_z \vert_{\text{total}} \nonumber \\
&  \rightarrow
\lambda \frac{  \left[4 \left(  |\tilde \mu| -|\mu| \right) + \mathrm{i}\, \omega_1 \,
\ln \left(\frac{4\, | \mu|\,   |\tilde \mu| + 2\,\mathrm{i}\,\omega_1 
\left(  |\tilde \mu| -|\mu| \right)+\omega_1^2}
{ 4\,  | \mu|\,   |\tilde \mu|-2 \,\mathrm{i}\, \omega_1 \left(  |\tilde \mu| -|\mu| \right)
+\omega_1^2}\right)\right ]\sigma_z}
  {64\, \sqrt{3} \,\pi }\,,
\end{align}
which is finite and does not contain the UV cutoff anymore.
\begin{widetext}
This gives the correction:
\begin{align}
& \delta \chi^{123}_1(\mathrm{i} \,\omega_1, \mathrm{i} \, \omega_2) 
\nonumber \\
= & 
\frac{\lambda \,e_A^3 \left [ \omega_1^3 \left ( \omega_1+2 \,\omega_2 \right ) \ln \left(4 \,\mu^2
+ \omega_1^2\right)- \omega_2^3 \left(2 \,\omega_1+\omega_2 \right ) \ln \left(4 \,\mu^2+\omega_2^2\right)
+\left (\omega_2-\omega_1 \right) \left(\omega_1+\omega_2 \right )^3 \ln \left(4 \,\mu^2+(\omega_1+\omega_2)^2\right) \right ]}
{1536 \,\sqrt{3}\, \pi ^3\,v\, \omega_1 \, \omega_2 \left (\omega_1+\omega_2 \right )} \nonumber \\
& \times   \left[4 \left(  |\tilde \mu| -|\mu| \right)
+ \mathrm{i} \left( \omega_1+\omega_2 \right)
\ln \left(\frac{ 4\,  | \mu|\,  |\tilde \mu| + 2\,\mathrm{i} \left (\omega_1+\omega_2 \right ) \left(  |\tilde \mu| -|\mu| \right)
+ \left( \omega_1+\omega_2 \right)^2}
{4\,  | \mu|\, \,  |\tilde \mu|-2 \,\mathrm{i}\left( \omega_1+\omega_2 \right) \left(  |\tilde \mu| -|\mu| \right)
+ \left( \omega_1+\omega_2 \right)^2}\right)\right ].
\end{align} 
\end{widetext}
Performing the analytical continuation $\mathrm{i}\, \omega_{1,2} \rightarrow \omega_{1,2} + \mathrm{i}\,\delta$, and setting $\omega_1 = - \omega_2 = \omega$, we find that this contributes as:
\begin{align}
& \delta \chi^{123}_1( \omega+\Omega, -\omega)
=  \delta \chi^{213}_1(  -\omega,\omega+\Omega) 
 \nonumber \\ & 
\overset{\Omega \rightarrow 0 } {=}
 \frac{ \lambda\,  e_A^3\, \omega^2 \left(  |\tilde \mu| -|\mu|\right)}
 { 192\, \sqrt{3} \,\pi ^2 \,v\, \Omega} 
 \, \Theta \big(\omega - 2\,|\mu| \big ) \,,
\end{align} 
which leads to the correction
\begin{align}
& \delta\left( \frac{d j_z} {dt} \right) \nn
& =-\frac{\mathrm{i}\,\lambda\, e_A^3\,\left(  |\tilde \mu| -|\mu|\right)}
 { 24 \,\sqrt{3}\,h^2 \,v} 
 \, \Theta \big(\omega - 2\,|\mu| \big )
 \left[ \mathbf{E}(\omega) \times \mathbf{E}(-\omega)  \right]_{z}  ,
\end{align} 
for the current in the $z-$direction.
Here, we have neglected the corrections to the chemical potentials, since they only change the frequency range within which CPGE for the the non-interacting case is nonzero. 

In a similar fashion, we get:
\begin{align}
& \delta \chi^{312}_1( \omega+\Omega, -\omega)=  \delta \chi^{132}_1( -\omega, \omega+\Omega)
\nonumber \\
 & \overset{\Omega \rightarrow 0 } {=}
 \frac{ \lambda \,e_A^3\, \omega^2 \left[ 4 \left( |\mu|- |\tilde \mu|\right)
-\omega \ln \Big |  \frac{(2 \,|\mu|+\omega ) (2\,|\tilde \mu|-\omega )}
  {(\omega -2 \,|  \mu| ) (2 \,|\tilde \mu|+\omega )} \Big |
 \right ] }
 { 768\, \sqrt{3} \,\pi ^2 \,v\, \Omega} \nonumber \\
& \hspace{ 1 cm} \times \Theta \big(\omega - 2\,|\mu| \big ) \,,
 \\
& \delta\left( \frac{d j_y} {dt} \right)  = \mathrm{i}\,\lambda\, e_A^3\,
\frac{ 4 \left( |\tilde \mu| -|\mu|\right)
+ \omega \ln \Big |  \frac{(2 \,|\mu|+\omega ) (2\,|\tilde \mu|-\omega )}
  {(\omega -2 \,|  \mu| ) (2 \,|\tilde \mu|+\omega )} \Big |}
 { 96 \,\sqrt{3}\,h^2 \,v} 
 \nonumber \\
 & \hspace{ 1.7 cm} \times \Theta \big(\omega - 2\,|\mu| \big )
 \left[ \mathbf{E}(\omega) \times \mathbf{E}(-\omega)  \right]_{y}   \,. 
\end{align} 
By symmetry in the $xy-$plane, we infer that 
\begin{align}
\delta\left( \frac{d j_x} {dt} \right)  &= \delta\left( \frac{d j_y} {dt} \right)   \,. 
\end{align} 
Due to the intrinsic anisotropy of the problem, it is not surprising that the corrections for the current
in the $z-$direction is different from that in the $xy-$plane.

\section{Summary and outlook}
We have computed the CPGE for the double-Weyl semimetal, first in the absence of interactions and then in the presence of short-ranged Hubbard interactions. In the non-interacting case, for low-enough frequency ranges of the applied electric field, the CPGE gets contribution only from one double-Weyl node and has a quantized value proportional to the topological charge of the corresponding node. However, switching on Hubbard interactions affects this result, destroying the quantization. This is similar to the results found for the case of CPGE currents in Weyl semimetals \cite{kozii}. The only difference is that the corrections for the current in the $z-$direction is different from that in the $xy-$plane, due to the anisotropic dispersion of the starting Hamiltonian. These results imply that unlike the quantum Hall effect in gapped phases or the chiral anomaly in field theories, the quantization of the CPGE in topological semimetals is not protected.

In future, it will be interesting to look at the corrections coming from the Coulomb interactions. The computations will be cumbersome for this case compared to the Weyl semimetal, due to the anisotropic dispersion of the double-Weyl Hamiltonian. It will also be interesting to see the effect of short-ranged correlated disorder on the CPGE, using the well-known techniques \cite{rahul-sid,ips-rahul,ips-qbt-sc}.  

\bibliography{biblio}

\end{document}